\documentstyle[11pt,epsf]{article}
 1
 1
 1

\renewcommand{\thefootnote}{\fnsymbol{footnote}}
\def\lsim{\mathrel{\raise.3ex\hbox{$<$\kern-.75em\lower1ex\hbox{$\sim$}}}}
\def\gsim{\mathrel{\raise.3ex\hbox{$>$\kern-.75em\lower1ex\hbox{$\sim$}}}}
\begin{document}
\noindent
\thispagestyle{empty}
\renewcommand{\thefootnote}{\fnsymbol{footnote}}
\begin{flushright}
\hfill
{\bf TTP96-19}\footnote[1]{The postscript file of this
preprint, including figures, is available via anonymous ftp at
ftp://www-ttp.physik.uni-karlsruhe.de (129.13.102.139) as 
/ttp96-19/ttp96-19.ps 
or via www at http://www-ttp.physik.uni-karlsruhe.de/cgi-bin/preprints/.}\\
{\bf hep-ph/9606288}\\
{\bf May 1996}\\
\end{flushright}
\vspace{3cm}
\begin{center}
  \begin{LARGE}
  \begin{bf}
TWO-LOOP CORRECTIONS TO FERMION PAIR \\[5mm]
PRODUCTION VERTICES\footnote[2]{
 Talk given at 31st Rencontres de Moriond: Electroweak Interactions and
Unified Theories, Les Arcs, France, 16-23 Mar 1996; to appear in the
proceedings.}
  \end{bf}
  \end{LARGE}
  \vspace{1.5cm}

\begin{large}
 A.H. Hoang
\end{large}
\begin{center}
   Institut f\"ur Theoretische Teilchenphysik,\\ 
   Universit\"at Karlsruhe, 
   D-76128 Karlsruhe, Germany\\  
\end{center}

  \vspace{4cm}
  {\bf Abstract}\\
\vspace{0.3cm}
\renewcommand{\thefootnote}{\arabic{footnote}}
\addtocounter{footnote}{-2}
\noindent
\begin{minipage}{12.0cm}
\begin{small}
Recent results on the entirely analytic
calculation of second order corrections to massive 
quark pair production vertices induced by light quark flavours are 
reviewed.
Based on the method presented in this talk the second order effects of
the insertion of any massless one-loop
vacuum polarization into the gluon line of
the first order diagrams can be determined. The behaviour of the
corrections at threshold for vector and axial-vector current induced
massive quark production is discussed.
\end{small}
\end{minipage}
\end{center}
\vspace{1.2cm}
\newpage
%
%
%
\noindent
\section{Introduction}
\label{sectionintroduction}
Within recent years a lot of work has been invested to determine 
two-loop radiative corrections to physical processes relevant for
present and future collider experiments. The aim of this difficult work
is twofold. On the one hand, one is looking for potentially large
corrections and on the other one likes to diminish the size of
the uncertainty of the corresponding one-loop calculations. Especially
in the framework of electroweak processes involving hadrons these
aspects are of importance, because the scale of the QCD coupling in the
one-loop corrections can only be fixed by an explicit two-loop
calculation.
\par
Of special interest is the process of hadron production in 
$e^+e^-$-collisions. Whereas the ${\cal{O}}(\alpha_s)$ corrections to
this process have been calculated quite a long time 
ago~\cite{KaeSab55, Zerwasaxial} for
all mass and energy assignments, the
complete ${\cal{O}}(\alpha_s^2)$ corrections have only been determined
as a high energy expansion up to terms of
${\cal{O}}(M^4/s^2)$~\cite{CheKueKwiRep}, 
where $\sqrt{s}$ denotes the c.m. energy
and $M$ the mass of the produced quarks. As far as $b\bar b$-production
at the $Z$ peak is concerned this is definitely a good approximation.
However, in view of forthcoming $e^+e^-$-collision experiments 
($\tau$-charm-, $B$-factory, NLC), where quark pairs will be produced
close to the threshold, the knowledge of the entire 
${\cal{O}}(\alpha_s^2)$ corrections for all values of $M^2/s$ is
mandatory.
\par
In our group two strategies are pursued to reach this aim. The first one
consists of finding a numerical approximation by use of the 
Pad\'e method. For that one takes information from the high energy
regime, the threshold region and at momentum transfer zero and
constructs an interpolating function based on conformal mapping of the
physical momentum range onto the unit circle in the complex plane.
The second strategy consists of the analytic calculation of the 
${\cal{O}}(\alpha_s^2)$ corrections. Of course this seems to be an
impossible task at the present stage. Especially certain classes of
gluonic corrections (those which do not exist in the corresponding QED
calculation) represent a great challenge. On the other hand, the
second order corrections, which arise from the
insertion of the vacuum polarization due to massless quarks
into the gluon line of the first order diagrams (belonging to the
so-called ``non-singlet'' corrections, see
Fig.~\ref{figferm}) are accessible. In particular, the corrections due to
massless quarks are sufficient to carry out the BLM
procedure~\cite{BrodskyBLM}, 
which sets the scale of the strong coupling in the
first order corrections and, if one believes in the
``naive non-abelianization'' hypothesis~\cite{BradhurstGroz}, 
even might provide good estimates for
the complete second order corrections.
\par
In this talk I will report on recent developments of the analytical
approach. I will concentrate on the calculation and the structure of the
light fermionic non-singlet second order corrections. 
Because the calculation is
most easily performed by using dispersion integration techniques, the
results are derived in on-shell renormalized QED. The transition to the 
$\overline{\mbox{MS}}$ scheme and to QCD is indicated afterwards. The
presentation  is kept general and can be applied to fermion anti-fermion
pair production vertices induced by any current. In particular, I will
discuss the threshold behaviour and the associated scales of quark pair 
production via the vector
and the axial-vector current. I will not talk about massive quark pair 
production mediated by singlet (or "double triangle") diagrams.
For a presentation of the Pad\'e method the interested reader is
referred to the talk by M.~Steinhauser~\cite{Steinhauser} 
and the references given therein.
\section{Calculation of the Light Fermionic Second Order Corrections}
\label{sectioncalculation}
\begin{figure}[ht]
 \begin{center}
 \begin{tabular}{ccc}
   \epsfxsize=3.5cm
   \leavevmode
   \epsffile[170 270 420 520]{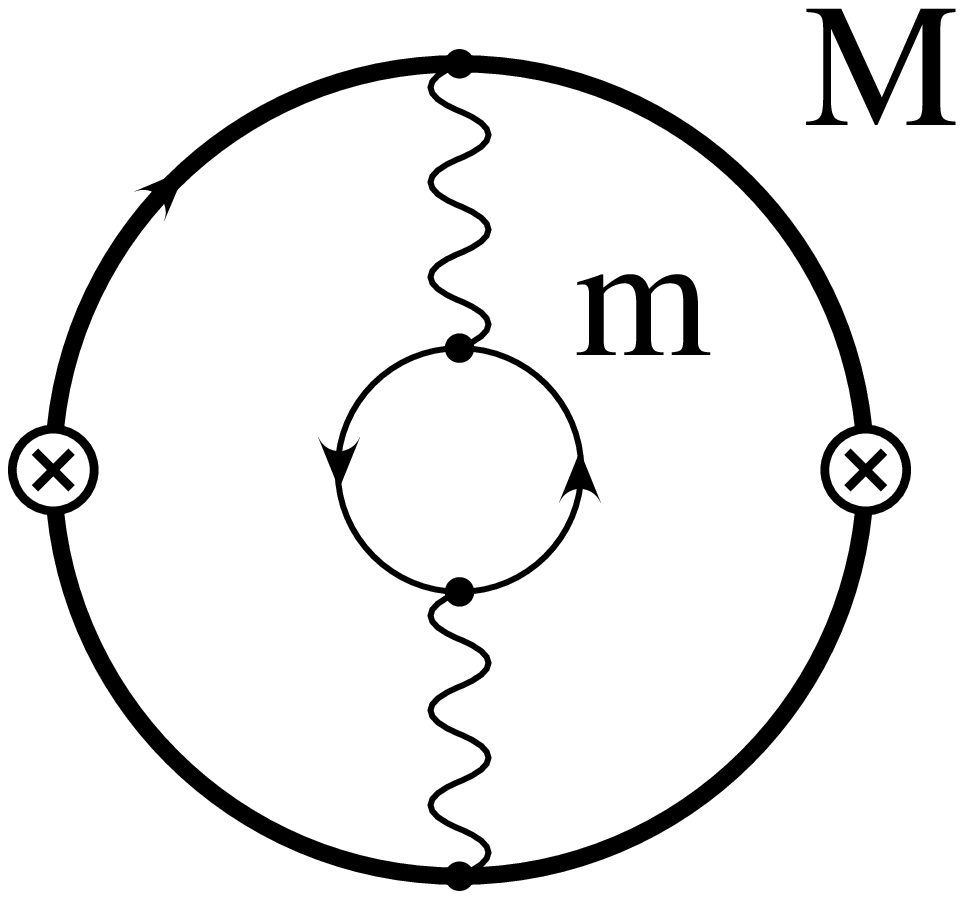}
   & \hspace{10ex} &
   \epsfxsize=3.5cm
   \leavevmode
   \epsffile[170 270 420 520]{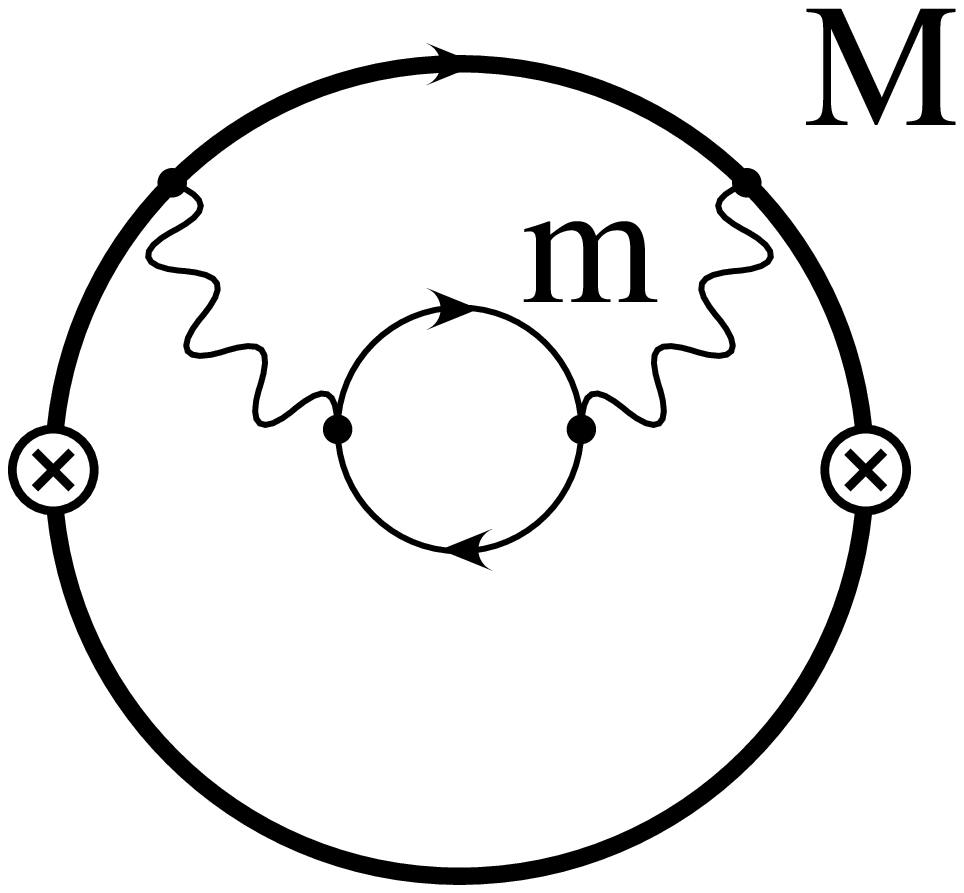}
 \end{tabular}
 \caption{\label{figferm} Non-singlet diagrams describing primary
production of massive fermion pairs (with fermion mass $M$) 
with additional real or virtual radiation of a light fermion 
anti-fermion pair (with fermion mass $m$).}
 \end{center}
\end{figure}

The non-singlet light fermionic second order corrections to the 
fermion anti-fermion pair production rate
\begin{equation}
R_{F\bar F}\, = \, r^{(0)} + 
               \,\bigg(\frac{\alpha}{\pi}\bigg)\,r^{(1)} +
               \,\bigg(\frac{\alpha}{\pi}\bigg)^2\,r^{(2)} +\, ...\,,
\label{RFFdef}
\end{equation}
correspond to the sum of all possible cuts of the diagrams depicted in
Fig.~\ref{figferm}.\footnote[1]{
The reverse situation with primary production of light fermions and
secondary radiation of massive ones has been treated in an earlier
work~\cite{HoaKueTeu94} and shall not be discussed here. 
Like the singlet
contribution they need not to be taken into account for the BLM
procedure as they do not contribute to the running of the coupling.
}
In the frame of on-shell renormalized QED the basic idea to calculate
those diagrams is to write the one-loop corrections to the photon
propagator in terms of a dispersion relation
\begin{eqnarray}
\lefteqn{
\frac{-i}{p^2+i\epsilon}
\left(g^{\mu\sigma}-\xi\frac{p^\mu\,p^\sigma}{p^2}\right)
\Big(-\,\Pi_{\sigma\rho}(m^2, p^2)\Big)
  \frac{-i}{p^2+i\epsilon} 
\left(g^{\rho\nu}-\xi\frac{p^\rho\,p^\nu}{p^2}\right)
}\nonumber\\
 & = &
 \frac{1}{3}\left(\frac{\alpha}{\pi}\right) \int\limits_{4\,m^2}^\infty
   \frac{{\rm d}\lambda^2}{\lambda^2}\,
   \bigg[\frac{-\,i\,g^{\mu\nu}}{p^2-\lambda^2+i\epsilon}\bigg]\,
   R_{f\bar f}(\lambda^2)\, + \ldots
\,\,,
\label{effectivepropagator}
\end{eqnarray}
where $R_{f\bar f}$ is the Born cross-section for $f\bar f$
production in $e^+e^-$-collisions normalized to the point cross-section.
It is evident that eq.~(\ref{effectivepropagator}) represents a
convolution of a massive vector boson propagator with the function 
$R_{f\bar f}$. The longitudinal and explicitly gauge dependent terms
indicated by the dots are irrelevant due to current conservation. Thus
the light fermionic second order corrections to $R_{F\bar F}$ can be
written as a one-dimensional integral
\begin{equation}
r_{f}^{(2)}(M^2, s) = 
 \frac{1}{3} \int\limits_{4\,m^2}^\infty
   \frac{{\rm d}\lambda^2}{\lambda^2}\, 
   r^{(1)}(M^2, s, \lambda^2)\,
   R_{f\bar f}(m^2, \lambda^2)\,,
\label{masterformulavectorvirtual}
\end{equation}
where $r^{(1)}(M^2, s, \lambda^2)$ represents the ${\cal{O}}(\alpha)$
corrections to the $F\bar F$ productions rate from the real and virtual
radiation of a vector boson with mass $\lambda$ at c.m. energy $\sqrt{s}$. 
For the special case of
interest, $m\to 0$, the evaluation of the 
integral~(\ref{masterformulavectorvirtual}) can be simplified enormously
by subtracting and adding $R_{f\bar f}$ at its asymptotic high energy
value $R^f_\infty\equiv R_{f\bar f}(\infty)$:
\begin{eqnarray}
r_{f}^{(2)}(M^2, s) & = &
 \frac{1}{3} \int\limits_{4\,m^2}^\infty
   \frac{{\rm d}\lambda^2}{\lambda^2}\, 
   r^{(1)}(M^2, s, \lambda^2)\,\left[
   R_{f\bar f}(m^2, \lambda^2)-R_\infty^f\right] 
\nonumber\\ & &
\,+\, \frac{1}{3}\,R_\infty^f \int\limits_{4\,m^2}^\infty
   \frac{{\rm d}\lambda^2}{\lambda^2}\, 
   r^{(1)}(M^2, s, \lambda^2)\,.
\label{rfermioniclight}
\end{eqnarray}
Because $R_{f\bar f}$ reaches its asymptotic value already for
$\lambda^2$ much smaller than the hard scales $M^2$ and $s$ we can
replace $r^{(1)}(M^2, s, \lambda^2)$ in the first integrand on 
the r.h.s. of eq.~(\ref{rfermioniclight}) by 
$r^{(1)}(M^2, s, 0)$, the one-loop contribution
to $R_{F\bar F}$ in eq.~(\ref{RFFdef}). 
The Bloch-Nordsieck theorem~\cite{BlochNord37}
assures that the corresponding limit exists.
The first integral can then be
trivially expressed in terms of the moment 
\begin{equation}
R_0^f \, \equiv \,
\int\limits_0^1 {\rm d}x\,
\Big[\,R_{f\bar f}\Big(\frac{4\,m^2}{x}\Big)-R_\infty^f\,\Big]
\,.
\label{momentsdef}
\end{equation}
The evaluation of the second
integral constitutes the main effort. The final result for the
second order corrections due to one light fermion $f$ can be 
written in the form ($x=f$)
\begin{eqnarray}
r_{x}^{(2)} & = &  
-\frac{1}{3}\Big[\, R_\infty^x\,\ln\frac{m^2}{s} - 
  R_0^x \,\Big]\, r^{(1)} +
  R_\infty^x\,\delta^{(2)}  
\,,
\label{finalresultonshell}
\end{eqnarray}
where $\delta^{(2)}$ is a complicated function of the ratio $M^2/s$
involving Tri- (${\rm Li}_3$), Di- (${\rm Li}_2$) and usual 
logarithms~\cite{ChHoaKueSteiTeu95,HTfuture}. 
For the light fermionic corrections we are interested in the moments
read
\begin{eqnarray}
R_\infty^f \, = \, 1\,, \qquad
R_0^f \, = \, \ln 4 - \frac{5}{3}\,\,.
\label{momentsfermions}
\end{eqnarray}
The result in the on-shell scheme can now be easily transferred to the 
$\overline{\mbox{MS}}$ scheme by expressing the fine structure constant
$\alpha$ in terms of the $\overline{\mbox{MS}}$ coupling
\begin{equation}
\alpha \, = \, \alpha_{\overline{\mbox{\tiny MS}}}(\mu^2) \,
           \left(\,1+\frac{\alpha_{\overline{\mbox{\tiny MS}}}(\mu^2)}{\pi}\,
                   \frac{1}{3}\,R_\infty^f\,\ln\frac{m^2}{\mu^2}\,\right)
          + {\cal O}(\alpha_{\overline{\mbox{\tiny MS}}}^3)
\,,
\label{alphaqedtorun}
\end{equation}
which effectively replaces $\ln(m^2/s)$ in~(\ref{finalresultonshell}) by
$\ln(\mu^2/s)$. At this point I would like to emphasize that the moments
can also be extracted directly from the vacuum polarization function due
to a massless fermion anti-fermion pair ($x=f$), 
\begin{eqnarray}
\Pi_{massless}^{x,\overline{\mbox{\tiny MS}}}(q^2)  
         & = &
 -\frac{\alpha_{\overline{\mbox{\tiny MS}}}(\mu^2)}{3\,\pi}\,\Big[\,
   R_\infty^{x}\,
   \ln\frac{-q^2}{4\,\mu^2} + R_0^{x}
   \,\Big]\,.
\label{vacpolmsbar}
\end{eqnarray}
\par
To arrive at the corresponding light fermionic second order corrections
in the frame of QCD we have to multiply the corresponding SU(3) group
theoretical factors $N_c=3$, $T=1/2$ and $C_F=4/3$. In particular the
moments for light quarks in QCD read
$R_n^{f,{\mbox{\tiny QCD}}} \, = \, T\, R_n^{f}$ ($n=\infty,0$).
\section{The Method of Moments}
\label{sectionmethodofmoments}
\begin{figure}[ht]
 \begin{center}
 \begin{tabular}{ccc}
   \epsfxsize=3.5cm
   \leavevmode
   \epsffile[170 270 420 520]{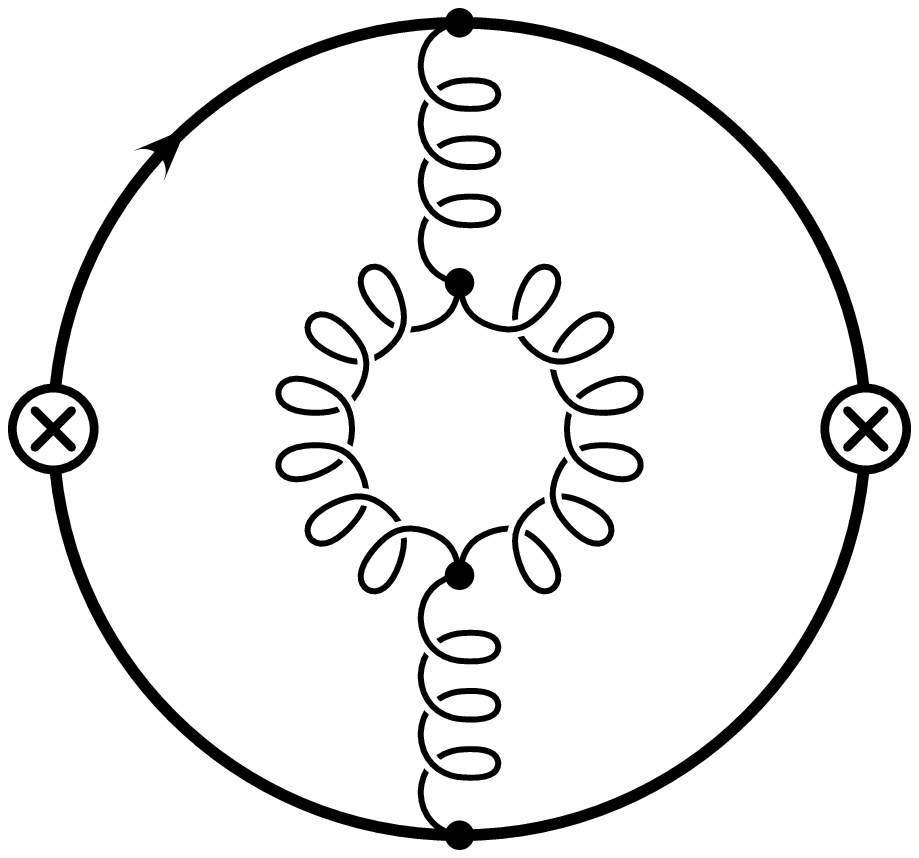}
   & \hspace{10ex} &
   \epsfxsize=3.5cm
   \leavevmode
   \epsffile[170 270 420 520]{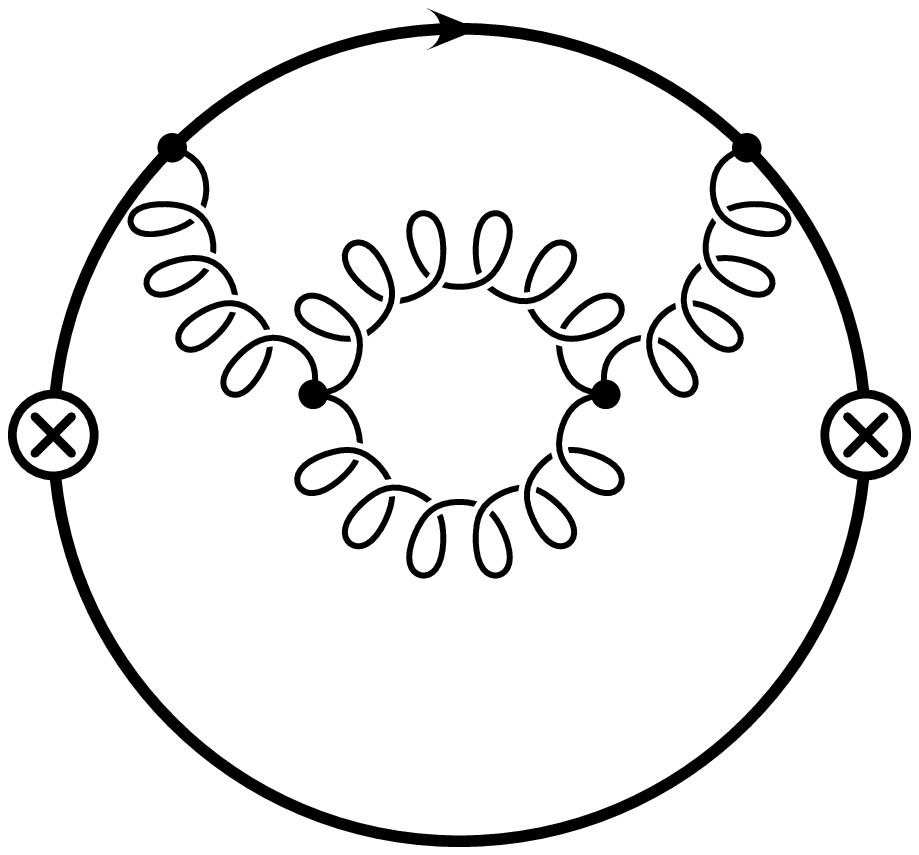}
 \end{tabular}
 \caption{\label{figgluons} Gluonic self-energy contribution to
heavy quark production. The ghost diagrams are not depicted explicitly.}
 \end{center}
\end{figure}

The method described above can be applied to
determine the corrections from the insertion of any massless vacuum
polarization into the gluon line, as long as the absorptive part of the
vacuum polarization  function approaches a constant in the high energy
limit. This is equivalent to the occurrence of at most one single power
of the logarithm $\ln(-q^2/\mu^2)$ in the renormalized vacuum
polarization function in $D=4$ dimensions and is certainly true for
any one-loop vacuum polarization function. Thus we can easily determine 
for example the second order corrections from the insertion of
a vacuum polarization due to massless coloured scalars into the gluon
line. According to eq.~(\ref{vacpolmsbar}) the scalar moments read
\begin{equation}
R_\infty^{s,{\mbox{\tiny QCD}}} \, = \, 
     T\,\left(\frac{1}{4}\right)\,, \qquad
R_0^{s,{\mbox{\tiny QCD}}} \, = \, 
     T\,\left(\frac{1}{4}\,\ln 4 - \frac{2}{3}\right)
\,.
\end{equation}
We are also in a position to calculate the second order
gluonic self-energy
contribution to massive quark production, as illustrated in 
Fig.~\ref{figgluons},
by determining the corresponding moments of the gluonic and ghost
contributions to the ${\cal{O}}(\alpha_s)$ gluon propagator
corrections,
\begin{eqnarray}
R_\infty^{g,{\mbox{\tiny QCD}}} 
& = & C_A\left(-\frac{5}{4} - \frac{3}{8}\xi\right),
\nonumber \\
R_0^{g,{\mbox{\tiny QCD}}} 
& = & C_A\left(\frac{31}{12}-\frac{3}{4}\xi+\frac{3}{16}\xi^2
                    +\left(-\frac{5}{4}-\frac{3}{8}\xi\right)\ln 4\right)
\,,
\label{gluonicmoments}
\end{eqnarray}
The gauge parameter $\xi$ is defined via the gluon propagator
in lowest order
\begin{equation}
\frac{i}{p^2+i\,\epsilon}\,\left(
-\,g^{\mu\nu} + \xi\,\frac{p^\mu\,p^\nu}{p^2}
\right)
\,.
\end{equation}
Evidently the gluonic moments and $r^{(2),{\mbox{\tiny QCD}}}_g$,
defined in analogy to eq.~(\ref{finalresultonshell}),
are not gauge invariant.
However, for $\xi=4$ the combination 
$\alpha_s^2\,R_\infty^{g,{\mbox{\tiny QCD}}}/(3\pi)$
coincides with the
complete gluonic ${\cal{O}}(\alpha_s^2)$ contribution to the QCD 
$\beta$-function, and thus 
$r_{g}^{(2),{\mbox{\tiny QCD}}}|_{\xi=4}$ accounts for the leading
logarithmic behaviour of the sum of all gluonic ${\cal{O}}(\alpha_s^2)$
diagrams in the high energy limit. It is not surprising that such a
gauge can be found.
However, it is a remarkable fact that for the same choice of $\xi$ also
the gluonic contributions to the perturbative QCD 
potential~\cite{Fischler77}
can solely be expressed in terms of the moments~(\ref{gluonicmoments}).
This point will be examined in more detail in the next section. Further
application of the method of moments to corrections to $R_{F\bar F}$
beyond the second order in the strong coupling can be found
in~\cite{ChHoaKueSteiTeu95,HTfuture}.
\section{The Threshold Behaviour}
\label{sectionthreshold}
The examination of the threshold region is of special interest. Here,
the contributions from the exchange of longitudinal and transversal
polarized gluons can be clearly distinguished. The former gives rise to
the instantaneous Coulomb interaction which cancels one power of 
$\beta=\sqrt{1-4\,M^2/s}$ from the phase space and leads to the 
well-known 
finiteness of the vector current cross-section at threshold. The
latter is responsible for the relativistic corrections. Both
contributions are governed by different momentum scales. From general
considerations one can infer that the scale relevant for the
instantaneous Coulomb interaction should be of the order of the relative
momentum of the produced massive quarks, $\sqrt{s}\,\beta$, whereas the
scale for the hard/transverse corrections should be of the order of the
c.m.~energy, $\sqrt{s}$. However, only an exact calculation of the 
non-singlet light
fermionic second order corrections allows for an accurate determination
of those scales via the BLM scale setting procedure.
\par
Taking in account only the second order effects of $n_f$ light quark 
species the threshold expansion of the massive quark pair production 
rate induced by
the vector current reads 
\begin{eqnarray}
\lefteqn{
R^V \, = \, 
N_c\,C_F\,\frac{3}{4}\,\alpha_s(\mu^2)\,\pi\,\bigg[\,
1+\frac{1}{3}\,\bigg(\frac{\alpha_s(\mu^2)}{\pi}\bigg)\,
n_f\,\bigg[\,
R_\infty^{f,\mbox{\tiny QCD}}\,\ln\frac{s\,\beta^2}{4\,\mu^2}
+R_0^{f,\mbox{\tiny QCD}}
\,\bigg]
\,\bigg] }
\nonumber\\ & & +\,
N_c\,\frac{3}{2}\,\bigg\{\,
1-4\,C_F\,\bigg(\frac{\alpha_s(\mu^2)}{\pi}\bigg)\,
\bigg[\,
1+\frac{1}{3}\,\bigg(\frac{\alpha_s(\mu^2)}{\pi}\bigg)\,
n_f\,\bigg[\,
R_\infty^{f,\mbox{\tiny QCD}}\,
\bigg(\,\ln\frac{s}{16\,\mu^2}+\frac{3}{4}
\,\bigg) 
+R_0^{f,\mbox{\tiny QCD}}
\,\bigg]
\,\bigg]
\,\bigg\}\,\beta 
\nonumber\\ & = & 
N_c\,C_F\,\frac{3}{4}\,\alpha_s(\mu^2)\,\pi\,\bigg[\,
1 - n_f\,\Pi_{massless}^{f,\overline{\mbox{\tiny MS}}}(-s\,\beta^2)
\,\bigg]
\nonumber\\ & &
+N_c\,\frac{3}{2}\,\bigg\{\,
1-4\,C_F\,\bigg(\frac{\alpha_s(\mu^2)}{\pi}\bigg)\,
\bigg[\,
1 - n_f\,\Pi_{massless}^{f,\overline{\mbox{\tiny MS}}}
    (-\frac{s}{4}\,e^{3/4})
\,\bigg]
\,\bigg\}\,\beta 
\label{RVthresh}
\end{eqnarray}
including terms up to ${\cal{O}}(\beta)$.
In the $V$-scheme, where the strong coupling is defined as the all order
effective charge in the QCD potential
\begin{eqnarray}
 V_{\mbox{\scriptsize QCD}}(Q^2) &=& 
         -4\pi \,C_F\,\frac{\alpha_V(Q^2)}{Q^2}
\,,
\label{QCDpotential}
\end{eqnarray}
with~\cite{Fischler77,Billoire80}
\begin{eqnarray}
 \alpha_V(Q^2) & = & 
\alpha_s(\mu^2)\,\bigg\{\,
      1 + \frac{\alpha_s(\mu^2)}{3\,\pi} \,
\bigg[\,
n_f \,T\,\bigg(  \ln\frac{Q^2}{\mu^2} - \frac{5}{3}
\,\bigg) \, + \,
C_A \,\bigg( 
          -\frac{11}{4}\ln\frac{Q^2}{\mu^2}+\frac{31}{12}
\bigg)
\,\bigg]
\,\bigg\} + {\cal{O}}(\alpha_s^3)
\nonumber \\ & = &
\alpha_s(\mu^2)\,
\bigg\{\,1
-n_f\,\Pi^{f,\overline{\mbox{\tiny MS}}}_{massless}(-Q^2)
-\Pi^{g,\overline{\mbox{\tiny MS}}}_{massless}(-Q^2)|_{\xi=4}
\,\bigg\} + {\cal{O}}(\alpha_s^3)
\,,
\label{alphav}
\end{eqnarray}
the corresponding BLM scales can be directly read off the arguments of
the vacuum polarization functions 
$\Pi^{f,\overline{\mbox{\tiny MS}}}_{massless}$ on the r.h.s. of
eq.~(\ref{RVthresh}). The reader should note that the second line of
eq.~(\ref{alphav}) is a new observation~\cite{ChHoaKueSteiTeu95} 
which has not been made in the original paper~\cite{Fischler77}.
Obviously the leading second order threshold contribution to
$R^V$ from massless quarks is in one-to-one correspondence to the form of
the QCD potential. This is in accordance with non-relativistic quantum 
mechanics.
As a consequence the leading second order gluonic threshold 
contributions to $R^V$, which are 
proportional to the colour factor $C_A$ (and represent purely 
non-abelian contributions) can be inferred direcly by adding 
$\Pi^{g,\overline{\mbox{\tiny MS}}}_{massless}(s\,\beta^2)|_{\xi=4}$ 
on the r.h.s. of eq.~(\ref{RVthresh}). A
similar conclusion is impossible for the hard corrections, because
they represent effects which cannot be described by the instantaneous 
Coulomb potential.
\par
Taking into account only the second order non-singlet
effects of $n_f$ light quarks
flavours the axial-vector induced massive quark pair production rate 
at threshold can be cast into form
\begin{eqnarray}
R^A & = & 
N_c\,C_F\,\frac{1}{2}\,\alpha_s(\mu^2)\,\pi\,\bigg[\,
1 - \Pi_{massless}^{x,\overline{\mbox{\tiny MS}}}(-s\,\beta^2\,e^{-2})
\,\bigg]\,\beta^2
\nonumber\\ & &
+N_c\,\bigg\{\,
1-2\,C_F\,\bigg(\frac{\alpha}{\pi}\bigg)\,
\bigg[\,
1 - \Pi_{massless}^{x,\overline{\mbox{\tiny MS}}}(-\frac{s}{4}\,e^{1/2})
\,\bigg]
\,\bigg\}\,\beta^3 + {\cal{O}}(\beta^4)
\,.
\label{RAVthresh}
\end{eqnarray}
The reader should note that the leading contribution to $R^A$ is of 
${\cal{O}}(\beta^2)$. In analogy to the vector current case the BLM
scales in the $V$-scheme can be identified from eq.~(\ref{RAVthresh}).
The ${\cal{O}}(\beta^2)$ second order gluonic corrections $\propto C_A$ 
can be determined in complete analogy to the vector current case. 
A formal proof of this statement will be presented in later work. 
\section{Acknowledgments}
\label{sectionacknowldgements}
I would like to thank my collaborators K.G. Chetyrkin, J.H. K\"uhn,
M. Steinhauser and T. Teubner for their important input to this work.
I also thank the {\it Graduiertenkolleg 
Ele\-men\-tar\-teil\-chen\-phy\-sik}
of the University Karlsruhe for financial
support of my work. Last but not least I thank
the organisers of {\it'96 Rencontres De Moriond}
for the very nice and inspiring time I had during my stay in Arcs~1800.

\sloppy
\raggedright
\def\app#1#2#3{{\it Act. Phys. Pol. }{\bf B #1} (#2) #3}
\def\apa#1#2#3{{\it Act. Phys. Austr.}{\bf #1} (#2) #3}
\def\lhc{Proc. LHC Workshop, CERN 90-10}
\def\npb#1#2#3{{\it Nucl. Phys. }{\bf B #1} (#2) #3}
\def\plb#1#2#3{{\it Phys. Lett. }{\bf B #1} (#2) #3}
\def\prd#1#2#3{{\it Phys. Rev. }{\bf D #1} (#2) #3}
\def\pR#1#2#3{{\it Phys. Rev. }{\bf #1} (#2) #3}
\def\prl#1#2#3{{\it Phys. Rev. Lett. }{\bf #1} (#2) #3}
\def\prc#1#2#3{{\it Phys. Reports }{\bf #1} (#2) #3}
\def\cpc#1#2#3{{\it Comp. Phys. Commun. }{\bf #1} (#2) #3}
\def\nim#1#2#3{{\it Nucl. Inst. Meth. }{\bf #1} (#2) #3}
\def\pr#1#2#3{{\it Phys. Reports }{\bf #1} (#2) #3}
\def\sovnp#1#2#3{{\it Sov. J. Nucl. Phys. }{\bf #1} (#2) #3}
\def\jl#1#2#3{{\it JETP Lett. }{\bf #1} (#2) #3}
\def\jet#1#2#3{{\it JETP Lett. }{\bf #1} (#2) #3}
\def\zpc#1#2#3{{\it Z. Phys. }{\bf C #1} (#2) #3}
\def\ptp#1#2#3{{\it Prog.~Theor.~Phys.~}{\bf #1} (#2) #3}
\def\nca#1#2#3{{\it Nouvo~Cim.~}{\bf #1A} (#2) #3}

\end{document}